\documentclass[a4paper,11pt]{article}
\pdfoutput=1 

\usepackage{jcappub} 

\usepackage[T1]{fontenc} 

\title{\boldmath Bose-Einstein condensate \& degenerate Fermi cored dark matter halos}

\author{W.-J. Chung}
\author{and L.A. Nelson}
\affiliation{Department of Physics \& Astronomy, Bishop's University\\2600 College, Sherbrooke  J1M 1Z7, Canada}

\emailAdd{wchung13@ubishops.ca}
\emailAdd{lnelson@ubishops.ca}

\abstract{There has been considerable interest in the last several years in support of the idea that galaxies and clusters could have highly condensed cores of dark matter (DM) within their central regions. In particular, it has been suggested that dark matter could form Bose--Einstein condensates (BECs) or degenerate Fermi cores. We examine these possibilities under the assumption that the core consists of highly condensed DM (either bosons or fermions) that is embedded in a diffuse envelope (e.g., isothermal sphere). The novelty of our approach is that we invoke composite polytropes to model spherical collisionless structures in a way that is physically intuitive and can be generalized to include other equations of state (EOSs). Our model is very amenable to the analysis of BEC cores (composed of ultra-light bosons) that have been proposed to resolve small-scale CDM anomalies. We show that the analysis can readily be applied to bosons with or without small repulsive self-interactions. With respect to degenerate Fermi cores, we confirm that fermionic particle masses between 1 -- 1000 keV are not excluded by the observations. Finally, we note that this approach can be extended to include a wide range of EOSs in addition to multi-component collisionless systems.}

\keywords{dark matter; galaxies: rotation; condensation: Bose-Einstein; equation of state; dark matter: condensation; dark matter: halo; dark matter: mass; particle: interaction; structure; scattering length}

\begin{document}
\maketitle
\flushbottom

\section{Introduction}
\label{sec:intro}

The dark matter problem remains one of the greatest outstanding
issues of modern physics. Dark matter is thought to comprise 84.5\% of the total mass
and 26.8\% of the total content of the Universe according to the latest measurements of
the Cosmic Microwave Background (CMB) by Planck\cite{PLANCK}. It is also
generally thought to play a key role in large scale cosmological structure formation and
galaxy evolution\cite{KSS}. The prevailing conjecture is that DM is composed of subatomic weakly-interacting massive particles (WIMPs) like supersymmetric neutralinos, axions, and sterile neutrinos, but experiments to detect them have failed to turn up conclusive evidence
of their existence\cite{Feng,Adikhari,Dev}. The only property known for certain about DM is that it
interacts gravitationally with baryonic matter and itself, but otherwise has little,
if any, significant interaction via the other fundamental forces. It is also possible that dark matter does not exist and that
the laws of gravitation must be modified (e.g., Modified Newtonian
Dynamics [MOND] and other alternate theories of gravity).

The most convincing evidence for dark matter (DM) is directly related to
its gravitational effects. This can be traced back to Zwicky's observations\cite{Ostriker,svdb,Zwicky}.  Later observations by Rubin 
showed that stars in the outer parts of spiral galaxies orbited with approximately the same rotational speed, a fact which
was irreconcilable with the visible baryonic matter distribution\cite{svdb,Rubin}. Astronomers soon proposed that the ``rotation curve problem''  
could be resolved if there was a roughly spherical halo of galactic dark matter.  Models of these halos range in complexity from the singular isothermal sphere halo model to the NFW (Navarro--Frenk--White) and Einasto profiles\cite{NFW,Einasto-1}. Some density profiles such as 
those of NFW and Einasto are constructed empirically based on N-body simulations or rotation curve data, while those of models such as the singular 
isothermal sphere are derived based on the assumed intrinsic properties of the non-interacting, dark-matter particles. 
The singular isothermal and NFW profiles have densities that diverge at the center of the halo (i.e., they are cuspy) while the
Einasto profile is designed to avoid this problem. 

The spherical nature of halos suggests that they may be modeled by polytropes (or combinations of polytropes). Polytropes were devised as spherically symmetric, self-gravitating
gaseous spheres that are solutions of the Lane-Emden equation where the pressure
of the gas (polytropic fluid) is related to its density through a so-called polytropic
equation of state (EOS) $P=K\rho^{1+\frac{1}{n}}$\cite{Kippenhahn}. The hydrodynamic structure 
of polytropes is completely determined by their polytropic index, $n$, 
and exhibit a universal scaling (i.e., are homologous for the same index). They can be excellent facsimiles of low-mass stars and white 
dwarfs\cite{Eddington,Mestel,Chandrasekhar,Hendry}, and their
relative computational simplicity makes them widely used as models in various
astrophysical contexts\cite{Horedt,Ibata}.

In this paper, we show for the first time that composite polytropes can provide a robust representation of the structure of cored dark-matter halos and that their implementation is motivated by the dynamics of the dark matter particles themselves. We will first outline the method of constructing composite polytropes, defining matching conditions at the interface of two polytropes with different indices ($n$). The model will then be applied to two important cases: The first case corresponds to DM that consists of bosons that have formed a hypothetical Bose--Einstein condensate (BEC) in the highly dense core. Boehmer \& Harko\cite{Boehmer} have postulated a halo composed entirely of such a BEC, and have shown that an $n=1$ polytrope can describe a BEC with quartic nonlinearity\cite{Dev}. This is especially tractable because $n=1$ is one of only three values of $n$ for which the Lane--Emden equation has closed-form analytic solutions. By using an $n=1$ polytrope to represent the core and an $n\rightarrow\infty$ polytrope for the collisionless isothermal envelope, we are able to construct a self-consistent bosonic halo and use it to generate rotation curves. The second case for which we apply our model corresponds to fermionic dark matter. This dark matter can form a degenerate Fermi gas at sufficiently high densities and can thus be described by an $n=1.5$ polytrope. The envelope is once again taken to be an $n\rightarrow\infty$ collisionless isothermal polytrope. In each case, we generate rotation curves for different particle properties and thus obtain constraints on the properties of the DM (e.g., mass and temperature). Finally, we note that this model is completely general for all cases wherein the DM can be described by an analogous polytropic fluid of index $n$. In particular, we demonstrate the power of the method by applying it to the recently proposed `fuzzy' bosonic DM\cite{SCB,Mocz}.

\section{Polytropes}
\label{sec:poly}

Under the appropriate physical conditions, the properties of collisionless systems in dynamical equilibrium can be modeled by simple polytropes. Given a specific Hamiltonian, Liouville's theorem implies that the phase-space volume is time-invariant. Letting $f(\vec{r},\vec{v};t)$ denote the usual phase-space particle density, the collisionless Boltzmann equation (CBE) follows\cite{BT}:

\begin{equation}
\frac{df }{d t}=\frac{\partial f }{\partial t}+\frac{\vec{p}}{m}\nabla f-\nabla \Phi\cdot \nabla_v f=0, \label{eq:CBE}
\end{equation}
where $\Phi$ is the (smoothed) gravitational potential of the system and is independent of time. Invoking Jeans' Theorem and imposing a spherical potential (i.e., rotational invariance), the mass density of an isotropic system can be expressed as

\begin{equation}
\rho=4\pi m\int_{0}^{\infty}f(E_r)v^2 dv. \label{eq:isorho}
\end{equation}
In order to take advantage of the insight that polytropes can provide, we consider distribution functions of the form

\begin{equation}
f(E_r)=\left\{
\begin{array}{c l}	
     f_0E_r^{n-\frac{3}{2}}, & E_r>0\\
     0, & E_r\leq 0
\end{array}\right., \label{eq:dist}
\end{equation}
where $f_0>0$ is a constant, and $n$ is constrained by $1/2 < n < 5$ in order to ensure convergence.
Substituting \eqref{eq:dist} into \eqref{eq:isorho}, we obtain 

\begin{equation}
\rho=m\left[\frac{\left(2\pi\right)^{\frac{3}{2}}\left(n-\frac{3}{2}\right)!}{n!}f_{0} \right ]=c_n\Phi^n,\qquad n>\frac{1}{2}. \label{eq:rhophi}
\end{equation}
Further substitution into the spherically symmetric Poisson equation yields

\begin{equation}
\frac{1}{r^2}\frac{d}{dr}\left ( r^2 \frac{d\Phi}{dr} \right )=-4\pi G c_n \Phi^n. \label{eq:lee}
\end{equation}
This equation has the same form as the Lane-Emden equation that describes polytropic spheres (see \eqref{eq:LE} below).

Polytropes were originally modeled as self-gravitating spheres of matter obeying a polytropic fluid EOS where the pressure $P$ was related to the density $\rho$ in the form\cite{Kippenhahn}:
\begin{equation}
P=K\rho^{1 + 1/n},\label{eq:Pp}
\end{equation}
where $K$ is the polytropic constant (constant of proportionality), $n$ is the polytropic index (equivalent to the parameter introduced in \eqref{eq:LE}), and $\gamma=1+\frac{1}{n}$ is the adiabatic index. We introduce dimensionless scaling variables
\begin{alignat}{1}
\xi\equiv\frac{r}{\alpha_n},\: & \theta^n\equiv\frac{\rho}{\rho_{0}}\label{eq:xi}
\end{alignat}
such that $\alpha_n$ is the scale radius and $\rho_{0}$ is a reference density (usually evaluated at the center of the polytrope). Starting from Poisson's equation for a spherical self-gravitating polytropic fluid, the dimensionless Lane-Emden equation is
\begin{equation}
\frac{1}{\xi^{2}}\frac{d}{d\xi}\left(\xi^{2}\frac{d\theta}{d\xi}\right)=-
\theta^{n}.\label{eq:LE}
\end{equation}
The solution $\theta(\xi)$ is unique for each index $n$, and thus spherical polytropes of a particular index $n$ form a \emph{homologous}
group\cite{Kippenhahn}.

There are several values of $n$ that admit closed-form analytic solutions to \eqref{eq:LE}. These include $n=0$, which corresponds to a constant density throughout the interior; $n=1$, which has a finite radius (i.e., $\theta=0$ at $\xi=\pi$); and $n=5$ (also known as the Plummer sphere), which has an infinite radius\cite{Horedt}. All polytropes with $n \geq 5$ have an infinite radius and thus have an infinite mass\cite{Horedt}. In order to obtain physically useful results, a radius is usually chosen at which to truncate the polytrope, similar to the procedure applied to DM halos. The $n=1$ case is of particular interest because it has been successfully shown by Boehmer \& Harko to be satisfied by a BEC with quartic nonlinearity\cite{Boehmer}.

Additionally, there are several values of $n$ for which no analytic solutions exist but are physically interesting. Eddington applied the $n=3$ polytrope to create the first modern model for the Sun\cite{Eddington,Mestel}. The $n=1.5$ polytrope was used by Chandrasekhar to model the structure of zero-temperature white dwarfs whose hydrostatic support is primarily derived from electron degeneracy pressure\cite{Chandrasekhar}. We shall see later that if the DM particles are fermionic and sufficiently dense, the $n=1.5$ polytrope can provide a robust model for degenerate DM.

Finally, the case $n=\infty$ corresponds to an isothermal ideal gas whose pressure EOS is $P=k_B\rho T/m=K_i\rho$, where $k_B$ is the Boltzmann constant, $m$ is the mass of the constituent particles, and $T$ is the temperature of the fluid. The isothermal Lane--Emden equation can be expressed in a somewhat different form\cite{Kippenhahn}:
\begin{equation}
\frac{1}{s^{2}}\frac{d}{ds}\left(s^{2}\frac{d\psi}{ds}\right)=
e^{-\psi}.\label{eq:ILE}
\end{equation}
Here, $s$ and $\psi$ are analogous to $\xi$ and $\theta$ in the case of the usual Lane-Emden equation \eqref{eq:LE}, respectively. They are defined such that
\begin{alignat}{1}
s \equiv\frac{r}{\alpha_{i}},\; & e^{-\psi}\equiv\frac{\rho}{\rho_{0}}.\label{eq:s}
\end{alignat}
It can be readily shown that the singular isothermal sphere (SIS), which has the form $\rho\propto r^{-2}$, satisfies the isothermal Lane--Emden equation \eqref{eq:ILE}. It is an example of a singular, or ``cuspy'' solution to the Lane--Emden equation. We will only be concerned with non-cuspy solutions (i.e., those that are finite at $r=0$).

The mass of an isothermal polytropic sphere of radius $R$ can be written as  
\begin{equation}
M= 4\pi\alpha_i^3\rho_0\int_{0}^{s_R}e^{-\psi}s^2ds=4\pi\alpha_i^3\rho_0s^2\frac{d\psi}{ds}\bigg|_{s_R} \ ,\label{eq:sphere_iso}
\end{equation}
where $R=\alpha_i s_R$.  For an infinite-radius polytrope such as the isothermal one, the appropriate $s_R$ can be adjusted for an assumed halo mass. The rotation curve is easily derived using $v_{rot}=\sqrt{{GM(r)}/{r}}$.

An extreme example of a dark matter candidate particle that could be modeled with only the isothermal polytrope is the so-called ``erebon.'' Proposed by Penrose as a possible dark matter candidate as a consequence of conformal cyclic cosmology (CCC), the erebon is an exceedingly massive particle with a mass of approximately $10^{-8}$ kg, or $10^{27}$ eV/$c^2$\cite{Penrose}. This is on the order of the Planck mass, roughly the mass of a grain of sand and $10^{23}$--$10^{27}$ times greater than most WIMP candidates, which have masses on the eV or keV scale. Thus the erebon would essentially be a classical Boltzmann particle and behaves like an ideal gas due to the fact that it can only interact gravitationally with other particles, including other erebons. 

\begin{figure}
\includegraphics[width=1.00\textwidth]{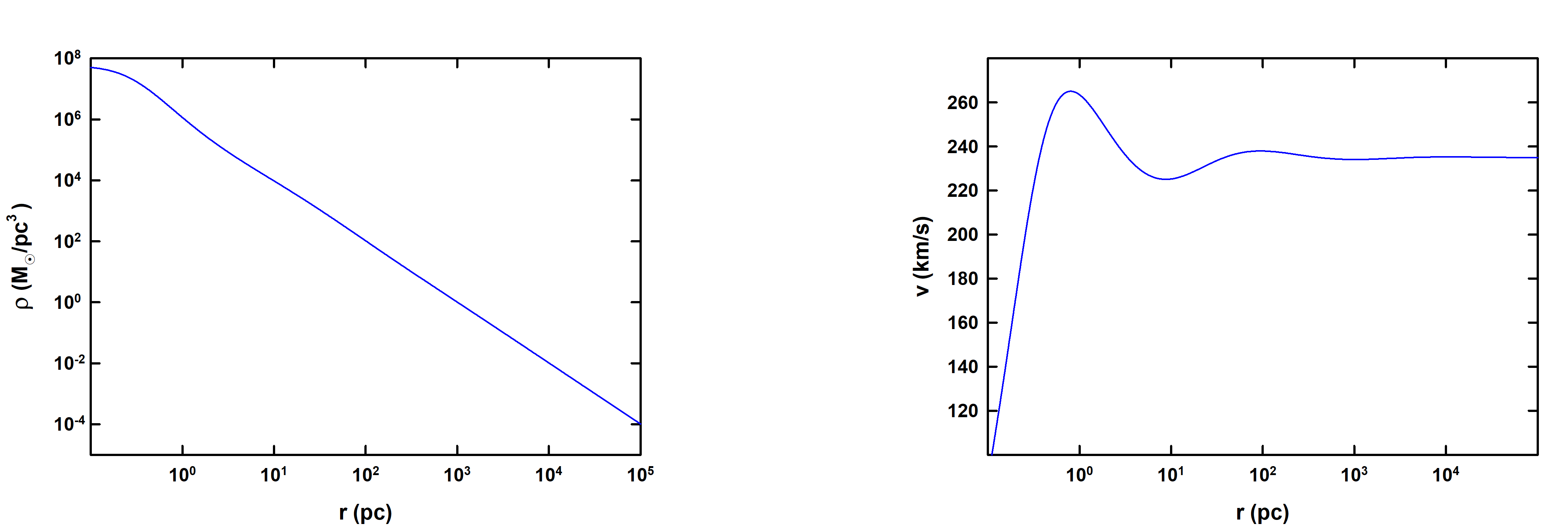}
\hfill
\caption{\label{fig:ere}Density profile for $10^{-8}$ kg erebons using
the isothermal polytrope model (left panel). The associated rotation
curve is shown in the right-hand panel.}
\end{figure}

Figure \ref{fig:ere} shows the density profile and rotation curve of a plausible galactic dark matter halo composed entirely of erebons. In order to obtain this profile, the temperature of the erebons was set to about $10^{26}$ K, about one millionth of the Planck temperature. This seems extremely hot, but by the equipartition of energy in an ideal gas, the average speed of the erebons is only about 200 km/s or $\sim0.001c$, safely in the non-relativistic regime. The total mass of the halo up to the cutoff radius of approximately $10^5$ pc is about $10^{12}~M_{\odot}$. Thus, it seems quite plausible that one can obtain a feasible dark matter halo even with such massive particles.

\section{Composite Polytropes}
\label{sec:com}

A composite polytrope is constructed of two or more polytropes with differing indices $n$ corresponding to various equations of state (EOSs) within different layers of the fluid. Such models have been successfully employed in the analysis of the evolution of low-mass main-sequence stars and brown dwarfs by Rappaport et al. and Nelson et al.\cite{Rappaport,Nelson}.  In order to create such a model, it is necessary to set the appropriate matching conditions of the physical properties of the object being modeled at each interface. In Rappaport et al., the composite polytrope (consisting of an $n=1.5$ core and $n=3$ envelope) was required to have matching masses, radii, densities, and pressures at the interface located at $r_i$. The corresponding Lane-Emden equations for the core
and envelope are thus
\begin{alignat}{1}
\frac{1}{\xi_{c}^{2}}\frac{d}{d\xi_{c}}\left(\xi_{c}^{2}\frac{d\theta_{c}}{d\xi_{c}}\right) & =-\theta_{c}^{3},\label{eq:LEc(5.75)}\\
\frac{1}{\xi_{e}^{2}}\frac{d}{d\xi_{e}}\left(\xi_{e}^{2}\frac{d\theta_{e}}{d\xi_{e}}\right) & =-\theta_{e}^{3/2},\label{eq:LEe(5.76)}
\end{alignat}
where the subscripts $c$ and $e$ denote the core and envelope, respectively. The associated densities, pressures, and radii at the interface are:
\begin{equation}
\begin{array}{ccc}
\rho_{c,i}=\rho_{0}\theta_{c,i}^{3},\: & P_{c,i}=K_{c}\rho_{0}^{4/3}\theta_{c,i}^{4},\: & \alpha_{c}\xi_{c,i}=\end{array}\left(\frac{K_{c}}{\pi G}\right)^{1/2}\rho_{0}^{-1/3}\xi_{c,i},\label{eq:scalec}
\end{equation}
\begin{equation}
\begin{array}{ccc}
\rho_{e,i}=\rho_{i}\theta_{e,i}^{3/2},\: & P_{e,i}=K_{e}\rho_{i}^{5/3}\theta_{e,i}^{5/2},\: & \alpha_{e}\xi_{e,i}=\end{array}\left(\frac{5K_{e}}{8\pi G}\right)^{1/2}\rho_{i}^{-1/6}\xi_{e,i}.\label{eq:scalee}
\end{equation}
The mass contained within radius $r=\alpha\xi$ is then
\begin{equation}
m_{c}(\xi_{c})=-4\pi\left(\frac{K_{c}}{\pi G}\right)^{3/2}\xi_{c}^{2}\theta_{c}^{'},\label{eq:mc}
\end{equation}
\begin{equation}
m_{e}(\xi_{e})=-4\pi\left(\frac{5K_{e}}{8\pi G}\right)^{3/2}\rho_{i}^{1/2}\xi_{e}^{2}\theta_{e}^{'},\label{eq:me}
\end{equation}
where prime notation is used to denote a derivative with respect to
$\xi$. By setting the respective quantities from the core and envelope to be equal at $\xi_{c,i}$ and $\xi_{e,i}$, they thus imposed constraints on the values of the variables $\theta$ and $\xi$ at the interface, and these can be written as:
\begin{equation}
\frac{\xi_{c,i}\theta_{c,i}^{3}}{\theta_{c,i}^{'}}=\frac{\xi_{e,i}\theta_{e,i}^{3/2}}{\theta_{e,i}^{'}},\label{eq:int1}
\end{equation}
\begin{equation}
\frac{8}{5}\frac{\xi_{c,i}\theta_{c,i}^{'}}{\theta_{c,i}}=\frac{\xi_{e,i}\theta_{e,i}^{'}}{\theta_{e,i}}.\label{eq:int2}
\end{equation}
To this must be added the boundary conditions at the core's center:
\begin{equation}
\theta_{c}(0)=1,\:\theta_{c}^{'}(0)=0.\label{eq:theta(5.83)}
\end{equation}
The normalization for the envelope's value of $\theta_{e}$ can be chosen arbitrarily ($\theta_{e,i}=1$ is chosen for convenience). 

Our models for the DM halo are constructed in a similar way. One of the advantages of using polytropes is that they are straightforward to compute. In addition, this type of model gives us a rather intuitive picture of the physics that governs the structure dynamics of the DM halo, and is also completely general.  If the DM's properties can be expressed as, or be approximated by a polytropic EOS, our method should be fully applicable. Finally, our analysis need not be restricted to the DM halo; the structure of any object whose constituents obey EOSs similar to that of the DM considered in this paper can be analyzed in an analogous way (e.g., dark-matter stars).

\section{Bosonic Halo}
 
Consider the case where the DM consists of bosons (e.g., axions\cite{Kuster}). Many of the small-scale structure anomalies associated with CDM (e.g., `cuspiness') can be resolved if ultra-light bosons can form a BEC\cite{Suarez}. The properties of the bosonic DM are further characterized by whether self-interactions are present. We will assume that  small, repulsive self-interactions occur and that at sufficiently high density, the DM will form a BEC, which can be described by a nonlinear variant of the Schr$\ddot{\text o}$dinger equation known as the Gross-Pitaevskii equation (GPE)\cite{Goodman,Boehmer}. Specifically, we take
\begin{equation}
-\frac{\hbar^{2}}{2m}\nabla^{2}\psi+V_{ext}\psi+g\left|\psi\right|^{2}\psi=\mu\psi,\label{eq:GPE}
\end{equation}
where $m$ is the boson mass, $V_{ext}$ is some external potential, $\mu$ is the chemical potential, $g=4\pi\hbar^{2}a/m$ is the effective two-body interaction between bosons defined by
\begin{equation}
U_{eff}(\mathbf{\vec{r},\,\overrightarrow{r}'})=g\delta(\mathbf{\vec{r}-\mathbf{\overrightarrow{r}'}}),\label{eq:Ueff}
\end{equation}
and $a$ is the effective scattering length.
According to the GPE, $\psi$ is the wave function of the entire BEC, which consists of $N$ individual bosons in the same coherent state $\phi$:
\begin{equation}
\psi=\sqrt{N}\phi,\label{eq:psi}
\end{equation}
The wave function of the BEC has the normalization
\begin{equation}
n_b=\left|\psi\right|^{2}, \label{eq:nb}
\end{equation}
\begin{equation}
N=\int d\mathbf{\overrightarrow{r}}\left|\psi\right|^{2}, \label{eq:N}
\end{equation}
 where $n_b$ is the boson number density.

In order to use a polytrope to model the BEC, we require the EOS of the BEC to have the form of \eqref{eq:Pp}. We shall now argue that the BEC or dilute Bose gas in the Thomas--Fermi approximation does in fact follow an $n=1$ polytropic EOS. Within this approximation, the first kinetic energy term in the GPE \eqref{eq:GPE} can be neglected, leaving only
\begin{equation}
V_{ext}\psi+g\left|\psi\right|^{2}\psi=\mu\psi.\label{eq:TF}
\end{equation}
For zero external potentials, this simply becomes
\begin{equation}
g\left|\psi\right|^{2}\psi=\mu\psi.\label{eq:v0}
\end{equation}
Thus,
\begin{equation}
\mu=gn. \label{eq:mu}
\end{equation}

Consider the uniform BEC in a fixed volume $V$. In order to add one boson to the BEC, we must add some internal energy $U(N+1)-U(N)$. The definition of the chemical potential $\mu$ implies that
\begin{equation}
\mu=\left(\frac{\partial U}{\partial N}\right)_{V}\equiv U(N+1)-U(N)=gn=g\frac{N}{V}. \label{eq:mu1}
\end{equation}
Given this, we can then find the internal energy $U$ of the BEC:
\begin{equation}
U(V,\,N)=\intop dN\mu=\intop dNg\frac{N}{V}=\frac{gN^{2}}{2V}. \label{eq:U}
\end{equation}
The pressure EOS can be derived from the thermodynamic definition of the pressure:
\begin{equation}
P=-\left(\frac{\partial U}{\partial V}\right)_{N}=\frac{gN^{2}}{2V^{2}}=\frac{g}{2m^{2}}\rho^{2}, \label{eq:P}
\end{equation}
which has the correct form of the EOS of an $n=1$ polytrope \eqref{eq:Pp}.

We can now construct the composite polytrope model for the bosonic dark matter halo with a BEC core. A similar cored bosonic--halo model has previously been explored by Slepian \& Goodman\cite{SG} (hereafter referred to as S\&G). However, S\&G chose a somewhat different approach, deriving an approximate equation of state for the bosons and solving the equations of hydrostatic equilibrium ($dP/dr=-(GM/r^2)\rho$) with it to obtain density profiles and rotation curves\cite{SG}. In our polytropic model, we first have to find the interface conditions in the region where the composite polytrope goes from the BEC EOS to the ideal gas EOS as the DM becomes less dense. By analogy to \eqref{eq:scalec}, \eqref{eq:scalee}, \eqref{eq:mc}, and \eqref{eq:me}, we conclude that:

\begin{equation}
\begin{array}{ccc}
\rho_{c,i}=\rho_{0}\theta_{i},\: & P_{c,i}=K_{c}\rho_{0}^{2}\theta_{i}^{2},\: & \alpha_{c}\xi_{i}=\end{array}\left(\frac{K_{c}}{2 \pi G}\right)^{1/2}\xi_{i},\label{eq:scalecbe}
\end{equation}
\begin{equation}
\begin{array}{ccc}
\rho_{e,i}=\rho_{i}e^{-\psi_{i}},\: & P_{e,i}=K_{e}\rho_{i}e^{-\psi_{i}},\: 
& \alpha_{e}s_{i}=\end{array}\left(\frac{K_{e}}{4\pi G \rho_i}\right)^{1/2}s_{i}.\label{eq:scaleebe}
\end{equation}
\begin{equation}
m_{c}(\xi_{i})=-\left(2\frac{K_{c}}{G}\right)\xi_{i}^{2}\theta_{i}^{'},\label{eq:mcbe}
\end{equation}
\begin{equation}
m_{e}(s_{i})=\left(\frac{K_{e}}{G}\right) s_{i}^{2}\psi_{i}^{'},\label{eq:mebe}
\end{equation}
Here, $K_c=g/2m^2$ and $K_e=k_B T/m$, as can be seen from the equations of state of the BEC and isothermal ideal gas, respectively. Setting the respective quantities to be equal to each other at the interface, we can also freely choose $\rho_{e,i}=\rho_i=\rho_{c,i}$ and $\psi_i=0$ for simplicity. The resulting constraints (cf. \eqref{eq:int1} and \eqref{eq:int2}) are thus

\begin{equation}
\frac{\xi_{i}\theta_{i}}{\theta_{i}^{'}}=-\frac{s_{i}}{\psi_{i}^{'}},\label{eq:intbe1}
\end{equation}
\begin{equation}
\frac{\xi_{i}\theta_{i}^{'}}{\theta_{i}}=-\frac{1}{2} s_{i}\psi_{i}^{'}.\label{eq:intbe2}
\end{equation}

Furthermore, the value of $\theta_i$ is determined by the interface density, which is in turn given by the equality of the pressure on both sides of the interface:

\begin{equation}
\rho_i=\rho_0 \theta_i = \frac{K_e}{K_c}. \label{eq:rhoi}
\end{equation}
Finally, it is necessary to determine $\theta_{i}^{'}$. Because the Lane-Emden equation reduces to a spherical Bessel function of the first kind for the case of $n=1$:

\begin{equation}
\theta=j_0 (\xi) = \frac{\sin \xi}{\xi}.\label{eq:theta1}
\end{equation}
Because the BEC solution is straightforward, we can now easily solve \eqref{eq:intbe1} and \eqref{eq:intbe2} to obtain the necessary constraints on the composite polytrope model. For other values of $n$ with no closed-form analytic solutions, it is necessary to resort to an approximation method in order to estimate $\theta$ and $\theta^{'}$ near the surface of the polytrope (i.e., where $\rho \rightarrow 0$). This is especially important for the case of a Fermi degenerate gas (see the next section). Details describing this type of calculation can be found in Appendix A. 

\begin{figure}
\includegraphics[width=1.00\textwidth]{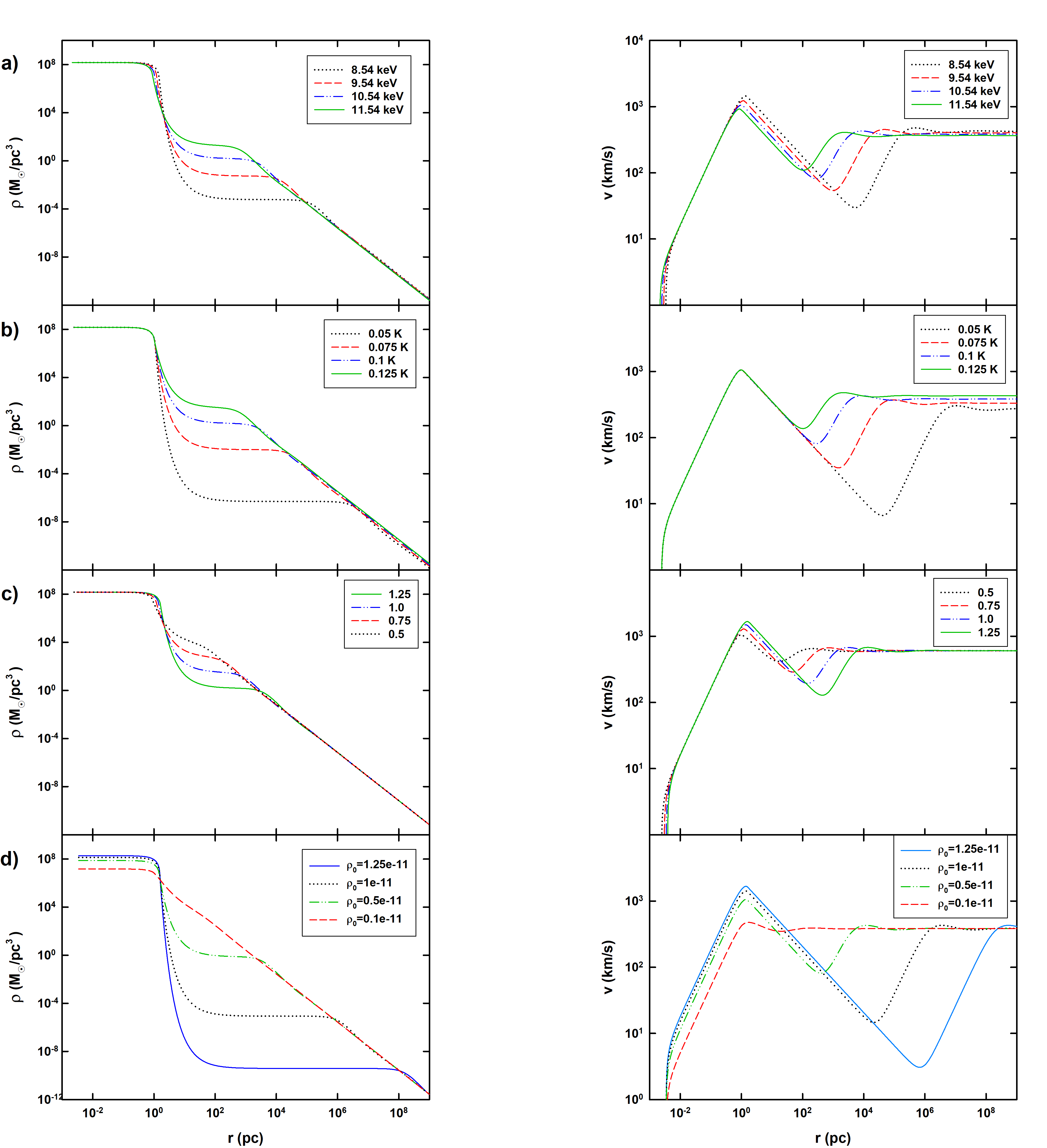}
\hfill
\caption{Density profiles (left-hand panels) and rotation curves (right-hand panels) for different bosonic dark-matter halo parameters: (a) particle mass, (b) temperature, (c) effective scattering length (in $10^{-14}$ m), and (d) central halo density (in units of kg m$^{-3}$).\label{fig:2}}
\end{figure}

We shall now explore how these BEC halos behave if certain parameters such as the temperature and the particle mass are varied. For example, in Figure \ref{fig:2}, we show how the density profile and rotation curve change as: (i) the mass of the particle; (ii) the temperature; (iii) the effective scattering length; and, (iv) the central density are varied.{\footnote {Because the Lane--Emden equation is a second-order, non-stiff, ordinary differential equation, the composite polytrope can be integrated numerically using the Runge--Kutta method. This method was used to generate all of the density and mass profiles, and the rotation curves presented for polytropes with non-closed-form solutions.}}

The left-hand panels of Figure \ref{fig:2} illustrate the density profiles and thus the structure of the composite BEC/isothermal halos. Of particular note are the ``plateaus'' in the density profiles, representing the transition regions going from BEC-like to ideal gas-like behavior. There are also very steep drops in the densities close to what would be the surfaces of the core ($n=1$) polytropes if the envelopes were removed. The polytropes' density profiles asymptotically approach those of isothermal spheres at large radii. This behavior ensures that flat rotation curves are obtained at large radii (as observed in numerous galaxies).{\footnote {Other density profiles such as $r^{-3}$ are admissible and our formulation is readily adaptable to those cases.}} One of the significant strengths of the composite polytrope model is that it can naturally blend the transition between the two regions in a smooth manner, in spite of the fact that the EOS abruptly changes from that of a BEC to an isothermal ideal gas at the interface.

The right-hand panels of Figure \ref{fig:2} show the associated rotation curves corresponding to the respective density profiles. We conclude that the isothermal envelope is much more sensitive to changes in the particle mass than the BEC core, which shows relatively little change with variations in the mass of the dark matter particle. There is also a sharp drop in the rotational velocity corresponding to the transition from the core to the envelope in each density profile. This seems to be a characteristic of cored composite polytrope halos because this feature is also present for the fermionic halos that are presented in the next section. It could therefore be a general indicator of cored structures in dark matter halos. Finally, it is worth noting that this general behaviour is consistent with that obtained by S\&G from their model using the approximate bosonic equation of state\cite{SG}.

While these results demonstrate that the composite polytrope model works well in reproducing previously computed results, it is important to note that observational constraints can be applied to these models. For example, S\&G have been able to show that a substantial portion of the parameter space quantifying $m$ and $a$ can likely be ruled out based on observations of the dynamics of the Bullet Cluster.  S\&G found that they could not obtain realistic rotation curves from their density profile unless their parameter $\theta$ was greater than $10^{-4}$. This led them to conclude that bosonic (repulsive) DM in thermal equilibrium is not viable. As for our models, it is possible that the bosons in the core and envelope have not thermalized because we are not assuming that the core and the envelope are in thermal equilibrium with each other. Nonetheless, this is an important issue that needs to be investigated further.{\footnote {We did find models that are not inconsistent with the observations if we set $m=10^{-4}$ eV, $\rho_0 = 2\times {10^{-17}}$ kg/m$^3$ (about 250 M$_\odot$/pc$^3$), $a = 10^{-25}$ m, $T=2\times 10^{-7}$ K). Note that the values of $a$ and $T$ are quite artificial.}}

\subsection{Fuzzy Bosonic Matter}

If the masses of the bosons are extremely light, bosonic quantum pressure (no self-interactions) is sufficient to prevent the DM halo from collapsing\cite{Hu}.  This type of bosonic DM halo problem has been proposed by Schive, Chiueh, and Broadhurst\cite{SCB} (hereafter SCB) and Mocz et al.\cite{Mocz}. They have proposed a wavelike DM ($\Psi$DM) that is composed of a BEC, thereby preventing a collapse on the Jeans' scale. According to their analysis, the structures of dwarf galaxies can have much shallower density profiles than would be predicted by CDM, thus more closely matching observations. The only free parameter in their theory is the bosonic mass $m_B$.

By solving the Schr\"odinger--Poisson equation numerically, SCB were able to calculate a solitonic density profile. Because of the scaling symmetry of the solitonic solution, SCB were able to characterize the density profile in terms of the core radius $r_c$ (where the density has dropped to 50\% of its maximum value). Based on this symmetry and from numerical results, SCB found that
\begin{equation}
\rho_c(r)\approx\frac{1.9m_B r_c^{-4}}{[1+0.091(r/r_c)^2]^8}\simeq 1.9 m_B r_c^{-4}[1 - 0.73(r/r_c)^2],~r\lesssim r_c,
\end{equation}
where $\rho_s$ has units of $M_\odot$/pc$^3$, $m_B$ is in units of $10^{-23}$ eV, and the radii are in kpc.

Taking $\rho_0=1.9 m_B r_c^{-4}$ and assuming $r/r_c \leq 1$, it can be shown that this closely matches the profile of an $n=2$ polytrope. The core polytrope then obeys the following EOS:
\begin{equation}
P_c=K_c \rho^{3/2},~K_c=C m_B^{1/2}.
\end{equation}
This core polytrope could then, in principle, be matched to an envelope polytrope with a transition region occuring near $r\geq r_c$. This $\Psi$DM model will be an intriguing application that we plan to explore further and is a prime example of the utility of the composite polytrope analysis.

\section{Fermionic Halo}

Many of the leading dark matter candidates are fermions such as neutralinos and sterile neutrinos\cite{Widrow,Bramante}. Unlike bosons, fermions obey the Pauli exclusion principle and will not form a Bose--Einstein condensate unless they condense as Cooper pairs. Instead, they can form a Fermi degenerate gas, and such a halo will primarily support itself from gravitational collapse through degeneracy pressure. We can conceive of such a DM halo as consisting of a degenerate core and an isothermal envelope.

Similar to the BEC halo investigated in the previous section, we will assume that the envelope is composed of isothermal fermions and that the EOS of the core corresponds to non-relativistic, degenerate fermions. The pressure EOS of the core is given by

\begin{equation}
P_{F}=\frac{1}{20}\left(\frac{3}{\pi}\right)^{2/3}\frac{h^{2}}{m_{F}{}^{8/3}}\rho^{5/3}=K_{F}\ \rho^{5/3} \label{eq:PF(5.84)}
\end{equation}
where we have assumed that the dark matter consists of only a single
species of fermion with mass $m_{F}$.{\footnote{This analysis can be extended to accommodate the co-existence of other fermionic species.}} This EOS corresponds to an $n=3/2$ polytrope with the Lane-Emden equation of the core given by
\begin{equation}
\frac{1}{\xi^{2}}\frac{d}{d\xi}\left(\xi^{2}\frac{d\theta}{d\xi}\right)=-\theta^{3/2}\ .\label{eq:LEc(5.86)}
\end{equation}
The associated core density, pressure, and scale length are
\begin{equation}
\begin{array}{ccc}
\rho_{c}=\rho_{0}\theta^{3/2},\: & P_{c}=K_{F}\rho_{0}^{5/3}\theta^{5/2},\: & \alpha_{c}=\end{array}\left(\frac{5K_{F}}{8\pi G}\right)^{1/2}\rho_{0}^{-1/6},\label{eq:scale(5.87)}
\end{equation}
where $\rho_{0}$ is the central density, $r_{c}=\alpha_{c}\xi$,
and the usual boundary conditions apply: $\theta(0)=1$, $\theta^{'}(0)=0$.
It is remarkable to note that the only free parameter describing the properties of the DM is its rest mass{\footnote {Technically $\rho_{0}$ is also a free parameter but its value is only needed to determine the physical properties (e.g., mass as a function of radius) inside the core.}}. The mass contained within radius $r_{c}$ is
\begin{equation}
m_{c}(\xi)=-4\pi\left(\frac{5K_{F}}{8\pi G}\right)^{3/2}\rho_{0}^{1/2}\xi^{2}\theta^{'}(\xi) \ . \label{eq:mc(5.88)}
\end{equation}

As for the isothermal envelope, we have again imposed the ideal gas EOS, and thus the corresponding equations from the previous section (\eqref{eq:scaleebe} and \eqref{eq:mebe}) apply 
here. We also introduce the new interface conditions, again requiring continuity of
radius, density, pressure, and mass (and also choosing $\psi_{i}=0$),
where the subscript $i$ denotes evaluation at the interface.
\begin{equation}
\frac{\xi_{i}\theta_{i}^{3/2}}{\theta_{i}^{'}}=-\frac{s_{i}}{\psi_{i}^{'}}\label{eq:int1(5.94)}
\end{equation}
\begin{equation}
\frac{\xi_{i}\theta_{i}^{'}}{\theta_{i}}=-\frac{2}{5}s_{i}\psi_{i}^{'}.\label{eq:int2(5.95)}
\end{equation}
The equality of pressures 
\begin{equation}
P_{c,i}=K_{F}\rho_{i}^{5/3}=P_{e,i}=K_{I}\rho_{i}.\label{eq:Peq)}
\end{equation}
implies that the interface density $\rho_{i}$ is independent
of the central density $\rho_{0}$, since $K_{F}$ and $K_{I}$ are
only dependent on the free parameters $m_{F}$ and $T$, respectively. Thus
\begin{equation}
\rho_{i}=\left(\frac{K_{I}}{K_{F}}\right)^{3/2}=\frac{(20k_{B}T)^{3/2}\pi m_{F}^{5/2}}{3h^{3}}=\frac{\sigma^{3}\pi20^{3/2}m_{F}^{4}}{3h^{3}}.\label{eq:rho_i(5.113)}
\end{equation}
We conclude that the specification of $\rho_{0}$, $K_{I}(T)$, and $K_{F}(m_{F})$ completely
specifies the structure of the composite polytrope{\footnote{To generate a physically reasonable composite model one should choose $\rho_{0}$ such that $\rho_{i}<\rho_{0}$.}}. 

\begin{figure}

\includegraphics[width=1.00\textwidth]{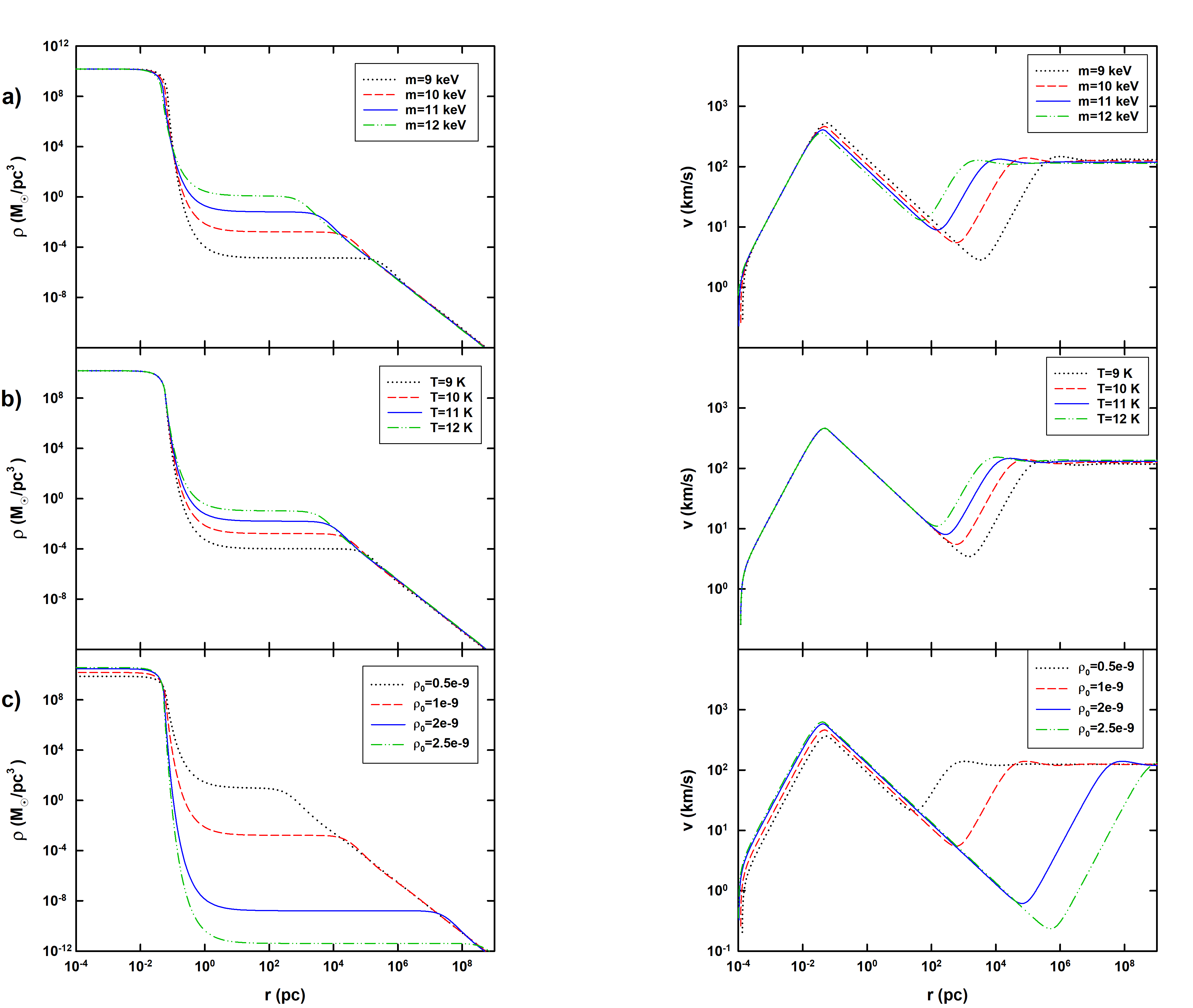}
\hfill

\protect\caption{Density profiles (left-hand panels) and rotation curves (right-hand panels) for different fermionic dark-matter halo parameters: (a) particle mass, (b) temperature, (c) effective scattering length, and (d) central halo density (in units of kg m$^{-3}$).\label{fig:6}}
\end{figure}

The composite polytrope model for a fermionic dark matter halo can be solved in the same way as for a bosonic halo. Figure \ref{fig:6} shows the density profiles and rotation curves of halos with different values for the mass, temperature, and central density parameters, respectively. It is important to note that the fermionic halos are less sensitive to changes in the respective parameters compared to the BEC ones.

It is most instructive to compare our approach to another scheme developed by Ruffini et al.\cite{RAR}. In the RAR model, Ruffini et al. also derive the density profile and associated rotation curve for a dark matter halo composed of massive fermions in a degenerate core, surrounded by a non-degenerate envelope obeying classical Boltzmann statistics. However, they used a rather different approach: they solved the general relativistic Einstein equations for a spherically
symmetric spacetime with the Tolman and Klein thermodynamic equilibrium equations. This is considerably more sophisticated than the polytropic approach that we are proposing, but it is similar in adopting a more ``bottom-up'' approach to creating a DM halo model, and thus the RAR model can be a useful check against our fermionic composite polytrope halo. 

\begin{figure}
\includegraphics[scale=0.09]{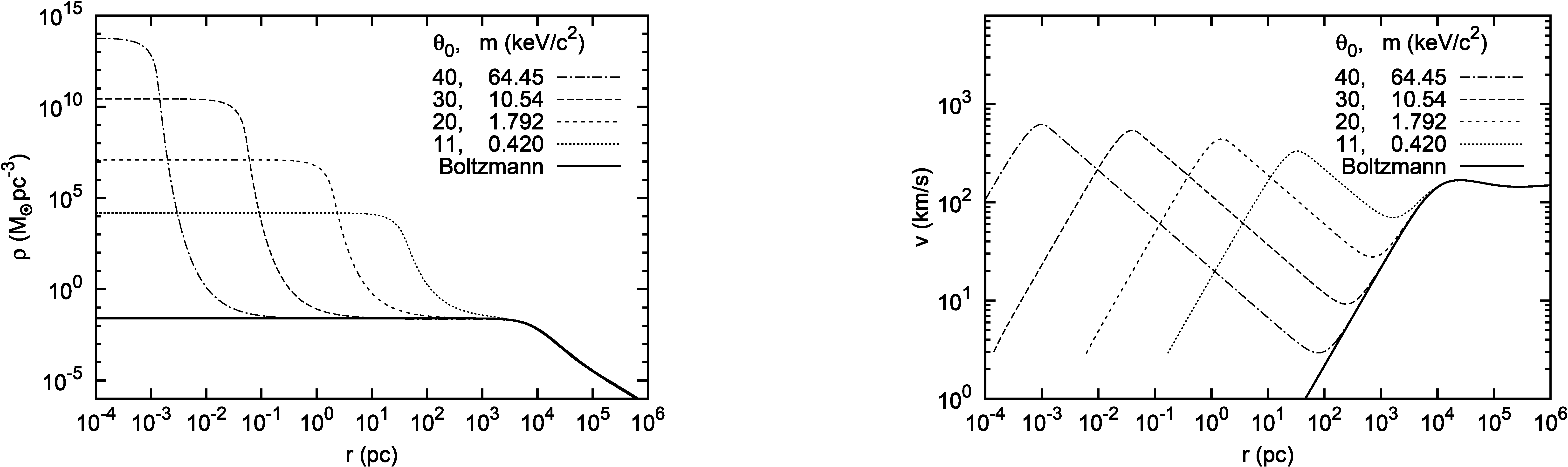}
\hfill

\protect\caption{The RAR model for the dark matter halo\cite{RAR}. In the left panel are the density
profiles, and on the right that are the corresponding rotation curves. $\theta_{0}$
is the central degeneracy parameter, which has no direct equivalent in our model but is roughly analogous to the central density. \label{fig:The-RAR-model}}
\end{figure}

\begin{figure}
\includegraphics[width=1.00\textwidth]{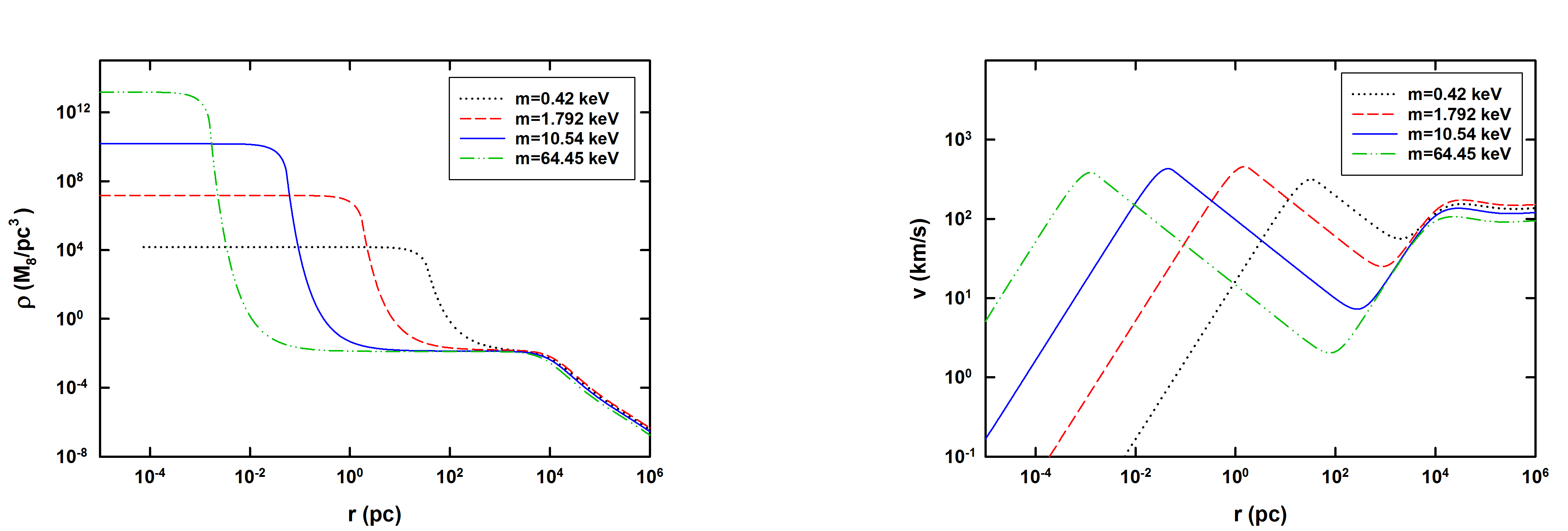}
\hfill

\protect\caption{Density profiles and rotation curves generated using the composite
polytrope approach with parameters similar to those used in the RAR model\cite{RAR} shown in Figure \ref{fig:The-RAR-model}. The
parameters ($m_{F}$, $T$, $\rho_{0}$) are: (64.45 keV, 37.5 K,
$1\times10^{-6}$kg/m$^{3}$), (10.54 keV, 10.0 K, $1\times10^{-9}$kg/m$^{3}$),
(1.792 keV, 2.75 K, $1\times10^{-12}$kg/m$^{3}$), (0.42 keV, 0.52
K, $1\times10^{-15}$kg/m$^{3}$).\label{fig:Density-profiles-and}}
\end{figure}

Figure \ref{fig:The-RAR-model} shows the density profile and rotation curve taken from \cite{RAR}.  Figure \ref{fig:Density-profiles-and} in turn shows a similar model using our composite polytrope. Since the parameters (other than the particle mass) in our model are not directly equivalent to those in the RAR model, they have been chosen to approximately match the same input physics. Our model reproduces the essential behavior of the RAR cored model with a high degree of fidelity. For instance, the density in the ``core'' region is relatively constant until the transition point is reached; this usually occurs near the point where one would expect the surface of the polytrope to be located (i.e., $\rho=0$, $P=0$) if it were an isolated polytrope.{\footnote{RAR have noted the very interesting implications arising from the fact that the masses of these degenerate cores are consistent with the masses of the putative intermediate and supermassive black holes thought to exist in different types of galaxies.}} Outside of the core, a transitional (intermediate) region exists where the DM starts to become isothermal. It is characterised by a steep drop in the density followed by a plateau as the isothermal EOS becomes dominant. Similar to the findings of RAR, we conclude that DM fermions with masses of between 1 to 1000 keV cannot be excluded by the observations.

Finally, in the fully isothermal regime at large radii, the density $\rho\sim r^{-2}$  (i.e., similar to that of the singular isothermal sphere). With a more fine-tuned choice of parameter values, the agreement between the two models would be even better. We also note that even though our polytropic models are completely Newtonian, we could refine our analysis to include relativistic polytropes (see, e.g., Demianski\cite{Demianski}). Nevertheless it is generally accepted that the inclusion of general relativity is not a particularly significant contribution to the dynamics of galactic DM halos; thus the Newtonian model should suffice. 

This agreement is quite remarkable even though our approach is motivated by the proven utility of using composite polytropes to model stars.{\footnote{Stars also exhibit abrupt transitions such as that which occurs near convective/radiative boundaries so the suitability for cored halos is not entirely unexpected.}}  Using \eqref{eq:sphere_iso}, the composite model allows us to conclude that the total mass of the halo can be expressed as
\begin{equation}
M=4\pi\left(\frac{K_{I}}{4\pi G}\right)^{3/2}\rho_{i}^{5/2}s_{f}^{2}\psi^{'}(s_{f}).\label{eq:M}
\end{equation}
where $s_f$ is the `cut-off' parameter.  At large radii, it can be shown from \eqref{eq:ILE} that $\psi^{'}_f \simeq 2/s_f$. Combining these two equations and using the definition $R=\alpha_e s_f$ yields
\begin{equation}
M=\frac{8\pi}{\alpha_e}\left(\frac{K_{I}}{4\pi G}\right)^{3/2}\rho_{i}^{5/2}R.\label{eq:M2}
\end{equation}
Once we substitute in $\rho_i$ from \eqref{eq:rho_i(5.113)}, we obtain a mass-radius relation such that $M=f(m_{F},\,T)R = f(m_{F},\,\sigma)R$.  The $M-R$ is \emph{independent} of the central density $\rho_{0}$ of the DM and only depends on the mass of the particle and the temperature of the halo (envelope).  Of course we have to specify $\rho_{0}$ to completely determine the structure and resulting rotation curve of the dark matter halo \emph{in the core}.  Another advantage of this approach is that we can straightforwardly derive other macroscopic relationships such as the $M-M_c$ relation that specifies how the mass of the halo and the degenerate core are related in terms of the microphysics (and temperature).  The dependence of the $M-M_c$ relation on the properties of the DM particles and the assumed envelope temperature, as well as the truncation radius, will be explored in a subsequent paper\cite{NC}.

\section{Conclusions}

Polytropic analysis is often the preferred method to gain physical insights into the structure and evolution of stars.  If the relationship between the momentum flux and distribution function of a collisionless system of DM particles obeys a polytropic EOS, this approach can also be used to analyze systems in approximate dynamical equilibrium.  We have shown that it is possible to build robust models of cored dark-matter halos using composite polytropes. This method is completely general and can be extended to core polytropes
with indices of $1/2 < n < 5$ (with an assumed isothermal envelope). 

To demonstrate this, we have explored two hypothetical halo models: one where the DM consists of self-interacting bosons forming a BEC that can be modeled by an $n=1$ polytrope \cite{Goodman,Boehmer}, and the other with fermions forming a degenerate core modeled by an $n=3/2$ polytrope \cite{RAR}.  The only properties assumed about the DM particles are that they obey Bose--Einstein or Fermi--Dirac statistics and behave like a collisionless ideal gas at sufficiently low densities. The viability of this method can be checked \emph{a posteriori} by comparing the concentration of particles ($n_d$) with the quantum concentration ($n_Q$).  We would expect $n_d \gtrsim n_Q$ in a degenerate core and $n \lesssim n_Q$ in the (isothermal) envelope.

For the range of parameters that we considered, the core of the DM halo is often many orders of magnitude smaller than the entire halo itself. The model shows that the central density of the halo has no bearing on the two principal macroscopic properties of the halo itself (i.e., its mass and radius); it only affects the properties within the core. It is the interface density $\rho_{i}$ that determines these macroscopic properties, and in turn, $\rho_{i}$ is constrained by the microscopic properties of the DM particles and the temperature/rotational velocity of the DM halo itself. This result only becomes readily apparent because of the physical tractability of the model, and it derives from the fact that the mass of the composite polytrope is required to be continuous across the interface. We note that use of the composite polytrope approach can be extended to include multi-component collisionless systems (see Horedt\cite{Horedt} for analogs).

We have also obtained a fairly simple relationship for the mass of the halo in terms of the cut-off radius. Any polytrope with index $n\geq5$
is unbounded (i.e., the density does not go to zero at some finite radius); therefore we must choose a radius at which to truncate the
polytrope in order to obtain a finite halo mass. This is not unique to the isothermal polytrope, and similar truncations must be done
when working with other halo models such as the NFW profile. However, this composite polytrope model allows us to link the microscopic properties of the DM particles to the macroscopic properties of the halo in a simple and physically intuitive way. This will be very beneficial in providing further constraints on the properties of DM particles based on astronomical observations\cite{NC}.

Finally, there remain some important questions to be explored: 1) Could a similar composite polytrope model also work for particles that do not follow the fermion or (self-interacting) boson EOSs?  We have shown that `fuzzy' DM halos seem to be tractable if the cores are modeled by $n=2$ polytropes.  Could more exotic EOSs, such as the strongly interacting massive particle (SIMP) EOS\cite{SIMP}, prove solvable using this approach? 2) What is the stability of the model to various types of perturbations? It is well known that polytropic gas spheres with indices $0 \leq n < 3$ are stable to small (radial) pressure perturbations far from the center, whereas $n>3$ polytropic and isothermal gas spheres are gravitationally unstable to such perturbations if they occur sufficiently far from the center\cite{Bonnor}. However, Bonnor also showed that incomplete (truncated) polytropes are stable to these perturbations if their radii are less than some critical radius determined by $n$\cite{Bonnor}.  It is not known whether such results are generally true for composite polytropes.  3) Can the method be extended to include the effects of rotation?   Generally this is only a perturbation for most cases but it can be studied using \emph{distorted polytropes}\cite{Horedt}. 4)  What are the implications with respect to astronomical observations?  Could these cores negate the need for supermassive (or intermediate-mass) black holes?  We plan to investigate all of these questions in detail.

\acknowledgments

We would like to thank the anonymous referee for a very informed and helpful report that greatly assisted us in improving the paper.  L.N. would like to thank the Natural Sciences and Engineering Research Council (Canada) for financial support provided through a Discovery grant. W.-J.C. acknowledges the Centre de Recherche en Astrophysique du Quebec (CRAQ) for graduate student support.  We also thank Calcul Qu\'ebec, the Canada Foundation for Innovation (CFI), NanoQu\'{e}bec, RMGA, and the Fonds de recherche du Qu\'{e}bec - Nature et technologies (FRQNT) for computational facilities.  Finally, we would also like to thank R. Dick, C.C. Dyer, V. Faraoni, C. Matzner, R.S. Orr, S. Rappaport for helpful discussions and M. Eby for technical assistance.

\appendix
\section{Surface Approximation}

If the interface occurs at $\theta\ll1$, i.e., very near
to the surface of the $n=1.5$ polytrope, it is possible to get an
analytical approximate solution for $\theta(\xi)$, and thus $\xi(\theta)$.
This was done by Chandrasekhar in 1939 using a Taylor series about
$\xi_{s}$\cite{Chandrasekhar-1}. The values of $\xi_{s}$ and $\theta_{s}^{'}$
(the surface values) are usually well-known for a bounded polytrope
from numerical integration, and Chandrasekhar was able to give a Taylor
series expansion in terms of these parameters and $\xi$:
\begin{equation}
\theta=-\xi_{s}\theta_{s}^{'}\left[\left(\frac{\xi_{s}-\xi}{\xi_{s}}\right)+\left(\frac{\xi_{s}-\xi}{\xi_{s}}\right)^{2}+\left(\frac{\xi_{s}-\xi}{\xi_{s}}\right)^{3}+...\right].\label{eq:theta(5.114)}
\end{equation}
The general Lane-Emden equation defining $\theta$ (see \eqref{eq:LE}) may be written
in the form
\begin{equation}
\theta^{''}+\frac{2}{\xi}\theta^{'}=-\theta^{n}.\label{eq:LEth(5.115)}
\end{equation}
By definition, $\theta_{s}\equiv\theta(\xi_{s})=0$, and thus near $\xi_{s}$ the Lane-Emden
equation is approximately ($n > 0$), 
\begin{equation}
\theta^{''}+\frac{2}{\xi}\theta^{'} \simeq 0.\label{eq:LEsur(5.116)}
\end{equation}
The solution to equation \eqref{eq:LEsur(5.116)} is given by 
\begin{equation}
\theta=-\xi_{s}\left|\theta_{s}^{'}\right|+\frac{\xi_{s}^{2}}{\xi}\left|\theta_{s}^{'}\right|.\label{eq:theta(5.117)}
\end{equation}
Thus, near $\xi_{s},$ one may approximate $\theta(\xi)$ with equation
\eqref{eq:theta(5.117)}. $\xi(\theta)$ is then given by
\begin{equation}
\xi(\theta)=\xi_{s}^{2}\left(\frac{\left|\theta_{s}^{'}\right|}{\theta+\xi_{s}\left|\theta_{s}^{'}\right|}\right).\label{eq:xi(5.118)}
\end{equation}
Since $\theta_{i}$ is determined by $K_{I}$, $K_{F}$, and $\rho_{0}$
(see equations \eqref{eq:rho_i(5.113)} and \eqref{eq:scale(5.87)}),
\eqref{eq:xi(5.118)} allows us to determine
$\xi_{i}$, provided that $\rho_{i}$ is sufficiently smaller than
$\rho_{0}$. Similarly, taking a derivative
with respect to $\xi$ also gives us an analytical approximation for
$\theta^{'}(\mbox{\ensuremath{\xi}})$ near $\theta=0$:
\begin{equation}
\theta^{'}(\xi)=-\left|\theta_{s}^{'}\right|\left(\frac{\xi_{s}}{\xi}\right)^{2}.\label{eq:theta'(5.119)}
\end{equation}
This is only a first-order approximation. It is possible to obtain
a better approximation that is valid\footnote{By valid, we mean to an accuracy in $\theta$ of 0.00025 at $0.5\xi_{s}$.}
at least as small as $0.5\xi_{s}$; such an
approximation was developed by Linnell\cite{Linnell}.
Linnell's approximation, however, has the same initial term \eqref{eq:theta(5.117)},
and it is the dominant term in the expansion in the same region\cite{Linnell}.
Thus, while \eqref{eq:theta(5.117)} may not be a very good approximation
well within the polytrope, it should remain a good approximation
as long as we are close to the surface of an isolated $n=1.5$ polytrope.


\end{document}